\documentclass[aps,prb,showpacs,twocolumn,floats,pdffig,pdflatex,floatfix]{revtex4}
\usepackage{amssymb}
\usepackage{amsbsy}
\usepackage{amsmath}
\usepackage{epsfig}
\usepackage{graphicx}
\usepackage{color}

\newcommand\beq{\begin{equation}}
\newcommand\eeq{\end{equation}}
\newcommand\bea{\begin{eqnarray}}
\newcommand\eea{\end{eqnarray}}

\begin{document}

\title {Josephson junctions detectors for Majorana modes and Dirac fermions}

\author{M. Maiti$^1$, K. M. Kulikov$^1$, K. Sengupta$^2$, and Y. M. Shukrinov$^1$}

\affiliation{$^1$ BLTP, JINR, Dubna, Moscow region, 141980, Russia.
\\ $^2$ Theoretical Physics Department, Indian Association for the
Cultivation of Science, Jadavpur, Kolkata 700 032, India.}
\date{\today}

\begin{abstract}

We demonstrate that the current-voltage (I-V) characteristics of
resistively and capacitively shunted Josephson junctions (RCSJ)
hosting localized subgap Majorana states provides a phase sensitive
method for their detection. The I-V characteristics of such RCSJs,
in contrast to their resistively shunted counterparts, exhibit
subharmonic odd Shapiro steps. These steps, owing to their
subharmonic nature, exhibit qualitatively different properties
compared to harmonic odd steps of conventional junctions. In
addition, the RCSJs hosting Majorana bound states also display an
additional sequence of steps  in the devil staircase structure seen
in their I-V characteristics; such sequence of steps make their I-V
characteristics qualitatively distinct from that of their
conventional counterparts. A similar study for RCSJs with graphene
superconducting junctions hosting Dirac-like quasiparticles reveals
that the Shapiro step width in their I-V curves bears a signature of
the transmission resonance phenomenon of their underlying Dirac
quasiparticles; consequently, these step widths exhibit a $\pi$
periodic oscillatory behavior with variation of the junction barrier
potential. We discuss experiments which can test our theory.
\end{abstract}


\maketitle

\section{Introduction}

The possibility of realization of Majorana zero modes, particles
with anyonic statistics described by real wavefunctions, has
attracted tremendous interest in recent years \cite{rev1}. Several
suggestions regarding condensed matter systems which can host such
fermions have recently been put forth
\cite{kit1,frac1,hel1,topo1,sup1,sup2,sup2a, sup3}. Out of these,
the most promising ones for experimental realization turns out to be
those which host Majorana modes as localized subgap states in their
superconducting ground state \cite{topo1,sup1,sup2,sup2a, sup3}.
Typically, the occurrence of such states require unconventional
superconducting pairing symmetry such as $p$- or $d$-wave
pairing\cite{pref,dref}. However recent proposals have circumvented
this requirement; it was shown that such bound states can occur
either at the end of one-dimensional (1D) wire in a magnetic field
with spin-orbit coupling and in the presence of a proximate $s$-wave
superconductor \cite{sup1,sup2} or in superconducting junctions atop
a topological insulator surface hosting Dirac fermions on the
surface of a topological insulators \cite{sup2a}. Such Majorana
fermions leave their signature as a midgap peak in tunneling
conductance measurement \cite{ks1} as well through fractional
Josephson effect \cite{ks2}.

Another interesting phenomenon in recent years has been the
discovery of materials whose low-energy quasiparticles obey
Dirac-like equations. These materials are commonly dubbed as Dirac
materials; graphene and topological insulators are common examples
of such materials \cite{grrev,tirev}. These materials can exhibit
superconductivity via proximity effect with Cooper pairing occurring
between Dirac electrons with opposite momentum \cite{been1,ks3}; it
is well known that transport properties of such superconductors
differ from their conventional counterparts and can serve as
experimental signatures of the Dirac nature of their constituent
quasiparticles \cite{been1,ks3}.

The experimental detection of Majorana modes have mainly relied on
measurement of either midgap peak \cite{exp1} or detection of even
Shapiro steps in a Josephson junctions of superconductors hosting
Majorana modes \cite{exp2}. The effectiveness of the former set of
experiments in detection of Majorana modes has been questioned since
the midgap peak did not lead to the expected $2e^2/h$ value of the
tunneling conductance and could have also occur from several other
effects such as presence of magnetic impurities leading to Kondo
effect \cite{kondoref1} and impurity induced subgap states
\cite{disref1}. In this sense, the presence of even Shapiro steps at
$V= n \hbar \omega_J/e$ (and the absence of odd ones at $(2n+1)\hbar
\omega_J/2e$) in Josephson current measurement, where $\omega_J$ is
the Josephson frequency, $n$ is an integer, and $V$ is the applied
external voltage, provide a more definite detection of such fermions
since they constitute a phase sensitive signature which is free of
effects of disorder \cite{ks2}. Consequently, theoretical studies of
AC Josephson effect for unconventional superconductors which hosts
Majorana modes has received a lot of attention lately
\cite{julia1,hassler1}. Theoretical studies of Josephson effect in
graphene Josephson junctions has also been carried out
\cite{been2,ks4}; it was shown that the critical current of such
junctions shows a novel oscillatory dependence on the barrier
potential of the junctions. However, the features of AC Josephson
effects in either of these systems for a resistively and
capacitively shunted Josephson junction (RCSJ) in the presence of an
external radiation has not been studied previously. In this context,
we mention that the analysis presented in the paper for junctions
hosting Majorana subgap states is somewhat idealized in the sense
that it does not provide a full treatment of Landau-Zenner tunneling
and other related quantum effects; however we do provide a
qualitative discussion and identify a regime of junction parameters
where our analysis is expected to hold qualitatively.

In this work, we study the Josephson junction described by a RCSJ
model where the superconductors forming the junction either host
Majorana modes at the interface or constitutes Dirac-like
quasiparticles. In the former case, we show that in contrast to
their counterpart in conventional junctions, the I-V characteristics
display {\it subharmonic odd Shapiro steps} whose width vanishes for
resistive junctions (in the limit where the junction capacitance
approaches zero) \cite{julia1,hassler1}. We provide an analytical
formula for the step width of both even and odd steps for such
junctions, show that the analytic result matches exact numerics
closely, and demonstrate, on basis of this analytical result, that
the behavior of this ratio is qualitatively different for Josephson
junctions with and without Majorana bound states. In particular we
demonstrate that the ratio of the odd and the even step width
decrease exponentially with the junction capacitance for junctions
with Majorana modes; in contrast, this ratio does not vary
appreciably for conventional $s$-wave junctions. Thus it serves as a
robust indicator for bound Majorana states in a JJ. Our analytical
results, supported by numerical analysis, reproduces the phenomenon
of absence of odd Shapiro steps in Josephson junctions with Majorana
bound states as a special limiting property of resistive Josephson
junctions; thus our work indicates that the absence of odd Shapiro
steps, while sufficient, is not a necessary characteristics of
Majorana bound states in a Josephson junction.  We also find that
the I-V characteristics of junctions with Majorana bound states show
a qualitatively different devil staircase structure which is
distinct from their $s$-wave counterparts. In particular, they
display additional sequence of steps which follow Farey's sum rule
\cite{fareyref}; such sequences are absent in I-V characteristics of
conventional $s$-wave junctions. In the latter case, for junctions
of superconductors hosting Dirac quasiparticles, we show that the
width of the Shapiro steps display $\pi$ periodic oscillatory
dependence on the barrier potential of the junction. Such a behavior
is a direct consequence of the transmission resonance phenomenon of
the Dirac-like quasiparticles of the superconductors forming the
junctions and is qualitatively distinct from conventional junctions
hosting quasiparticles which obey Schrodinger equation. We note that
our work shows that a RCSJ can act as phase sensitive detection
device for both Majorana bound states in topological superconductors
and Dirac like quasiparticles in a superconductor; it is therefore
expected to be of interest to theorists and experimentalists working
on both Majorana modes and Dirac materials.

The plan of the rest of the work is as follows. In Sec.\ \ref{maj1},
we provide our analytical and numerical results for junctions which
hosts subgap Majorana bound states. This is followed by Sec.\
\ref{dir1}, where we present our results on junctions which hosts
Dirac-like quasiparticles. Finally, we summarize our main results,
provide a discussion of experiments that can test our theory, and
conclude in Sec.\ \ref{conc1}.

\section{ Junctions with Majorana modes}
\label{maj1}

\begin{figure}[t!]
\includegraphics[width=0.9\columnwidth]{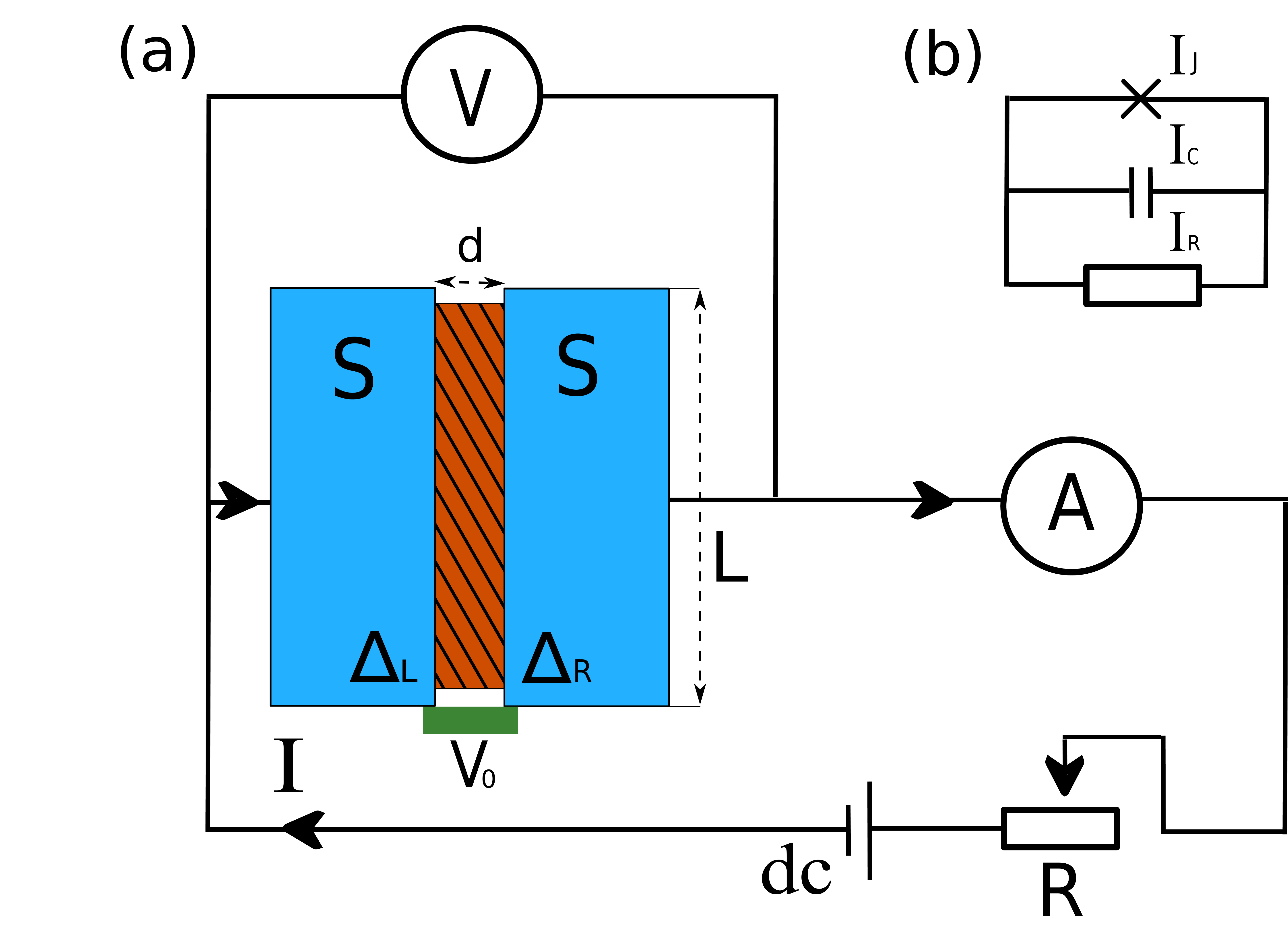}
\caption{(Color online) Schematic representation of the Josephson
junction in the RCSJ circuit (see inset). The junction has width $L$
in the transverse direction and the barrier region separating the
two superconductors has a thickness $d$ and is modeled by a
potential $V_0$. $I_J$, $I_C$ and $I_R$ are the Josephson current,
displacement current and quasiparticle currents respectively.}
\label{fig1}
\end{figure}

The basic design of the circuit which we propose to serve as
detector is shown in Fig.\ \ref{fig1}. To analyze the property of
this circuit, we first consider the Josephson junction. In our
proposal, this comprises of two superconductors with order
parameters $\Delta_R$ (for $x>d/2$) and $\Delta_L$ (for $x<-d/2$)
separated by a barrier region of width $d$ ($-d/2 \le x \le d/2$)
characterized by a barrier potential $V_0$ as shown in Fig
\ref{fig1}. The superconductors can either be topological
superconductors with (effective) $p$-wave pairing \cite{sup1,sup2}
or superconductors with Dirac-like quasiparticles which has $s$-wave
pair-potential \cite{been1,been2,ks3,ks4}. In this section, we
analyze the former case in details. We note at the outset that the
present analysis will hold for topological superconductors in 1D
wire geometry \cite{sup1,sup2} provided that the transverse
dimension $L$ is set to zero.

Josephson junctions shown in Fig.\ \ref{fig1}, are known to support
localized subgap Andreev bound states which can be obtained as
solution of the Bogoliubov-de Gennes (BdG) equation. For topological
superconductors which support $p$-wave pairing, the BdG equations
reads
\begin{eqnarray}
\left[(H_{\beta} +V(x)) \tau_z + (\Delta_{\beta}(x) \tau_{+} + {\rm
h.c.}) \right] \psi_{\beta} = E \psi_{\beta} \label{bdg1}
\end{eqnarray}
where $\beta= R,L$ for the right and the left superconductors,
$\psi_{\beta}= [\psi_{\beta \uparrow}(x, k_{\parallel}), \psi_{\beta
\downarrow}^{\dagger}(x,k_{\parallel})]$ is the two component BdG
wavefunction, $V(x)=V_0 \delta(x)$ is the barrier potential (we take
the limit of a thin barrier for which $d\to 0$), $H_{\beta}= \hbar^2
k^2/(2m) - \mu$ denotes the dispersion of the left and right
superconductors with $\mu$ being the chemical potential, $m$ the
electron mass, and $k^2= -\partial_x^2 + k_{\parallel}^2$. In what
follows, we are going to assume $p_x$-wave pairing and write
$\Delta_L(x)= \Delta_0 k_{xF}/k_{F}$ and $\Delta_R = \Delta_0
k_{xF}/k_{F} \exp(i \phi)$, where $k_{F}$ are the Fermi momenta of
the two superconductors, $\phi$ is the phase difference across the
junction, $\Delta_0$ is the amplitude of the superconducting gap,
and $k_{xF}$ is the $x$ component of the Fermi momentum. The
wavefunctions $\psi_{\beta}$ satisfies the boundary condition
$\psi_L(x=0)= \psi_R(x=0)$ and $\partial_x \psi_L(x=0) -
\partial_x \psi_R(x=0) = k_F \chi_1(k_{\parallel})\psi_L(x=0)$, where
$\chi_1(k_{\parallel})\equiv \chi_1= 2 U_0/\hbar v_F(k_{\parallel})$
is the dimensionless barrier potential for a given transverse
momentum of the quasiparticles. The localized subgap solutions of
Eq.\ \ref{bdg1} are given by \cite{ks2,comment1}
\begin{eqnarray}
E_1= -\Delta_0 \cos(\phi/2)/\sqrt{1+\chi_1^2/4}. \label{enmaj}
\end{eqnarray}
Note that for $\phi=\pi$, $E=0$ and it has been shown that this
state constitutes a realization of Majorana modes \cite{sup2a}.
Since these subgap states are the only ones with $\phi$ dependent
dispersion, one find the zero temperature Josephson current as
\begin{eqnarray}
I_1(\phi)= \frac{2e}{\hbar} \partial E_1/\partial \phi =
\int_{-k_F}^{k_F}  \frac{ d k_y}{2\pi} \frac{e \Delta_0
\sin(\phi/2)}{2 \hbar \sqrt{1+\chi_1^2/4}}. \label{jos1}
\end{eqnarray}
As noted in Ref.\ \cite{ks2}, the current is $4\pi$ periodic and a
substitution $\phi \to 2eVt/\hbar$ in the presence of a bias voltage
$V$ leads to fractional AC Josephson effect \cite{ks2,comment1}.

We now use Eqs.\ \ref{jos1} to obtain the response of an RCSJ
circuit constructed out of the superconducting junctions discussed
above. The RCSJ model, shown in the inset of Fig.\ \ref{fig1},
include a resistive component to take into account dissipative
process into account which may occur, for example, due to
quasiparticle tunneling and a shunting capacitance $C$ which takes
into account the displacement currents due to possible charge
accumulation in the leads \cite{rcsjref1}. The current
phase-relationship for this model, in the presence of an external
radiation, is given by \cite{rcsjref1,yury1}
\begin{eqnarray}
\ddot \phi + \beta \dot \phi + I_J(\phi)/I_c &=& I/I_c +
A\sin(\omega t)/I_c, \label{rcsj1}
\end{eqnarray}
where $I_c$ is the critical current of the junction, $A$ and
$\omega$ are the amplitude and frequency of the external radiation,
$I_J(\phi) = I_1$ for superconducting junctions with Majorana
fermions, $\beta= \sqrt{\hbar/(2e I_c R^2 C_0)}$, $R$ and $C_0$
denote the resistance and capacitance of the junction, and we have
scaled $t \to t/\tau$, where $\tau= \sqrt{\hbar C_0/(2 e I_c)}$.

Before proceeding further, we note that Eq.\ \ref{enmaj} holds in
the ideal limit where the two subgap Majorana modes at the two ends
of each wire do not interact \cite{hassler1}. Such an interaction
leads to hybridization of amplitude $\delta$ between the two
Majorana branches: $E_{1 \rm {hyb}} = \pm \sqrt{ \delta^2 + E_1^2}$
\cite{hassler1}. The hybridization amplitude $\delta$ arising from
such interaction is exponentially suppressed for long wires $\delta
\sim \exp[-L/(2\xi)]$, where the coherence length $\xi$, for 1D
nanowires, depends on the spin-orbit coupling strength of the wire.
In the presence of an external voltage, this leads to a
Landau-Zenner tunneling probability $P_{\rm t} = \exp[- 2 \pi
\delta^2/(E_J \hbar \dot \phi)]$, where $E_J \sim \Delta_0$ is the
maximal Josephson energy. The manifestation of $4 \pi$ periodicity
is evident when $P_{t} \simeq 1$. This occurs when $\delta^2/(E_J
\hbar \dot \phi) \ll 1$; a complete determination of frequency and
voltage range where $P_t \simeq 1$ requires the self-consistent
solution of Eq.\ \ref{rcsj1} and $I_J = \partial E_{2 {\rm
hyb}}/\partial \phi$. We have not attempted this in this work;
however, we note that such a regime can always be obtained for long
enough wire since $\delta$ is exponentially suppressed in this
regime leading to $P_t \simeq 1$. An analysis of these conditions
for resistive junctions can be found in Ref.\ \onlinecite{hassler1};
in the rest of this work, we shall assume $P_t \simeq 1$ and work
with Eq.\ \ref{enmaj}.

In what follows, we first obtain an approximate analytical solution
of Eq.\ \ref{rcsj1} in Sec.\ \ref{majan} which demonstrates the
existence of subharmonic odd Shapiro steps for junctions hosting
Majorana bound states. This is followed by Sec. \ \ref{majnu} where
we carry out a detailed numerical study of Eq.\ \ref{rcsj1} where
the analytical results are verified and the devil staircase
structure of the Shapiro steps is studied.

\subsection{Peturbative analytical Solution}
\label{majan}

In this section we consider perturbative analysis of Eq.\
\ref{rcsj1} for an unconventional Josephson junction comprised of
superconductors hosting subgap Majorana states with $E=E_J \cos(\phi
/2)$. We begin from the equation of such a junction given by Eq.\
\ref{rcsj1} and analyze this equation for $\omega,\beta \omega, A
>> 1$. The key point regarding this analysis is the observation in
the regime mentioned above it is possible to expend $\phi$ as
\cite{sub1}
\begin{eqnarray}
\phi &=& \sum_n \epsilon^n \phi_n , \quad  I = \sum_{n=0}^{\infty}
\epsilon^n I_n \label{expan1}
\end{eqnarray}
where $I_0$ is the applied current and $I_{n}$, $\epsilon \ll 1$,
and $I_n$ for $n>0$ determined self-consistently from the condition
of absence of additional dc voltage: $ \lim_{T \to \infty} \int_0^T
\dot \phi_n =0$ \cite{sub1}.

The equations for $\phi_n$ can be obtained by equating terms in the
same order of $\epsilon$. The procedure is standard and yields
\begin{eqnarray}
\ddot \phi_n + \beta \dot \phi_n &=& f_n(t) + I_n, \quad
f_0 = A \sin(\omega t),\nonumber\\
f_1 &=& -\sin[\phi_0/2], \quad f_2= \phi_1 \cos[\phi_0/2]/2
\label{epeqs}
\end{eqnarray}
Note that the $n=0$ equation represents the autonomous I-V curve of
the junction and is independent of the non-linear sinusoidal term.

To solve these equations, we note that this represent a linear first
order differential equation in $\dot \phi_n$; consequently we define
$y_n = \dot \phi_n $ and write
\begin{equation}
\dot y_n + \beta y_n = I_n + f_{n}(t). \label{yeq1}
\end{equation}
These equations admit a solution
\begin{eqnarray}
y_n(t) &=& I_n + e^{-\beta t} \int_{0}^{t} e^{\beta
t'}f_{n}(t')dt', \nonumber\\
\phi_{n}(t) &=& \int_{0}^{t} y_{n}(t')dt' + \phi_{n}(0).
\end{eqnarray}
For $n=0$, this yields
\begin{equation}
\phi_{0}(t) = \phi' + I_0 t + \frac{A}{\omega \gamma} \sin(\omega
t+\alpha_0), \label{4}
\end{equation}
where $\alpha_0=\arccos (\omega/\gamma)$, $\gamma = \sqrt{\beta^2+
\omega^2}$, and $\phi'$ is the DC phase of the junction. The
supercurrent at this order is given by
\begin{eqnarray}
I_{s}^{(0)}  &\sim& \sin(\phi_0 (t)/2) = {\rm Im} (e^{i\phi_0(t)/2})\\
&=&{\rm Im} \sum_{n=-\infty}^{\infty} J_n(x) e^{i([I_0/2+n\omega]t +
n\alpha_0 + \phi'/2)}, \label{supcurr}
\end{eqnarray}
where $x= A/(2 \gamma \omega)$. Thus the Shapiro steps occur when
the AC component of the supercurrent vanish: $I_0=2|n| \omega$ (even
steps). The width of the $n^{\rm th}$ even step can be read off from
Eq.\ \ref{supcurr}: $W_{\rm even} = \Delta I_{s}^{\rm even} =
2J_{n}(x)$. These yield expressions for the width of harmonic steps.
Note that in contrast to the conventional junctions where $I_s \sim
\sin[\phi_0]$, only even harmonic steps occur for junctions which
host subgap Majorana steps.

Next, we obtain the solution for $\phi_1$. Substituting Eq.\
\ref{supcurr} in Eq.\ \ref{yeq1}, we find after some straightforward
algebra
\begin{equation}
\phi_1 = \sum_{n=-\infty}^{\infty} J_n(x) (\gamma_n \omega_n)^{-1}
\cos(\omega_n t + n\alpha_0 + \delta_0 +n\phi'/2). \label{phieq}
\end{equation}
where $\omega_n=I_0/2+n\omega$, $\delta_n= \arccos(
\omega_n/\gamma_n)$, and $\gamma_n=\sqrt{\omega_n^2 + \beta^2}$. At
this order, the supercurrent is given by
\begin{eqnarray}
I_{s}^{(1)} &\sim& \frac{1}{2} \phi_1(t) \cos(\phi_0(t)/2)\nonumber\\
&=& \sum_{n_1,n_2 =-\infty}^{\infty} J_{n_1}(x) J_{n_2}(x)
(2 \gamma_{n_1} \omega_{n_1})^{-1} \nonumber\\
&& \times \sin(\omega_{n_1}t + n_1(\alpha_0+\phi'/2) + \delta_{n_1})
\\
&& \times \cos(\omega_{n_2}t +  n_2(\alpha_0+\phi'/2) + \delta_{n_2}) \nonumber\\
&=& \sum_{n_1,n_2} J_{n_1}(x) J_{n_2}(x) (4 \gamma_{n_1}
\omega_{n_1})^{-1}  \nonumber\\
&& \times \Big[\sin([\omega_{n_1} + \omega_{n_2}]t + [n_1 +
n_2](\alpha_0+\phi'/2) + \delta_{n_1}) \nonumber\\
&& + \sin([\omega_{n_1} - \omega_{n_2}]t + [n_1 -
n_2](\alpha_0+\phi'/2) + \delta_{n_1}) \Big] \nonumber\
\label{supcurr2}
\end{eqnarray}

At this order, we find that there are additional steps in the DC
component of the supercurrent; these steps occur at $|n_1 +n_2|
\omega=I_0$ for which the first of the two terms in the right side
of Eq.\ \ref{supcurr2} become independent of time. A set of these
steps occur at $(n_1+n_2)=2n-1$ for integers $n=1,2...$ and
constitute the odd Shapiro steps. Thus we find that the odd steps
for a junction of superconductors hosting Majorana ground states are
necessarily of subharmonic nature. The width of these steps can be
read off from Eq.\ \ref{supcurr2} to be
\begin{eqnarray}
W_{\rm odd} &=& \Delta I_{s n}^{\rm odd} =  \sum_{n_1}
\frac{J_{n_1}(x) J_{2n-1-n_1}(x)}{ 2 [(\{2n-1- n_1\}\omega)^2
+\beta^2]} \label{step1}
\end{eqnarray}
We note that when $C_0 \to 0$, $\beta \to \infty$ and the
subharmonic steps vanish leading to the result that only even
harmonic Shapiro steps exists for resistive Josephson junctions
hosting subgap Majorana steps. Thus our analysis reproduces the
absence of odd Shapiro steps in Josephson junctions with Majorana
bound states as a special case \cite{ks2,julia1,hassler1}. We also
note that these odd steps have a completely different origin than
the analogous steps discussed in Ref.\ \onlinecite{hassler1} since
they occur without any $\sim \sin(\phi)$ dependence of $I_J$.
Finally, we note that the ratio of the $n^{\rm th}$ even and the
adjacent odd Shapiro steps for these junctions are given by
\begin{eqnarray}
\eta_{n} &=& \frac{W_{\rm even}}{W_{\rm odd}} = \frac{2
J_{n}(x)}{\sum_{n_1} \frac{J_{n_1}(x) J_{2n+1-n_1}(x)}{ 2 [(\{2n-1-
n_1\}\omega)^2 +\beta^2]} } \label{etaan}
\end{eqnarray}
In the next section, we shall compare the analytical expression
(Eq.\ \ref{etaan}) with numerical results obtained by exact
numerical solution of Eq.\ \ref{rcsj1}.

\subsection{Exact numerical results}
\label{majnu}

To compute the I-V characteristics, we study the temporal dependence
of $V = \hbar \dot \phi/(2e)$ obtained by numerical solution of Eq.\
\ref{rcsj1} as a function of time for a fixed bias current $I$. The
dc component of the voltage is obtained by standard procedure
\cite{yury1,rcsj2} from $V$ and plotted as a function of $I$ to
generate the I-V characteristics.

\begin{figure}[t!]
\begin{center}
\includegraphics[width=0.49\columnwidth]{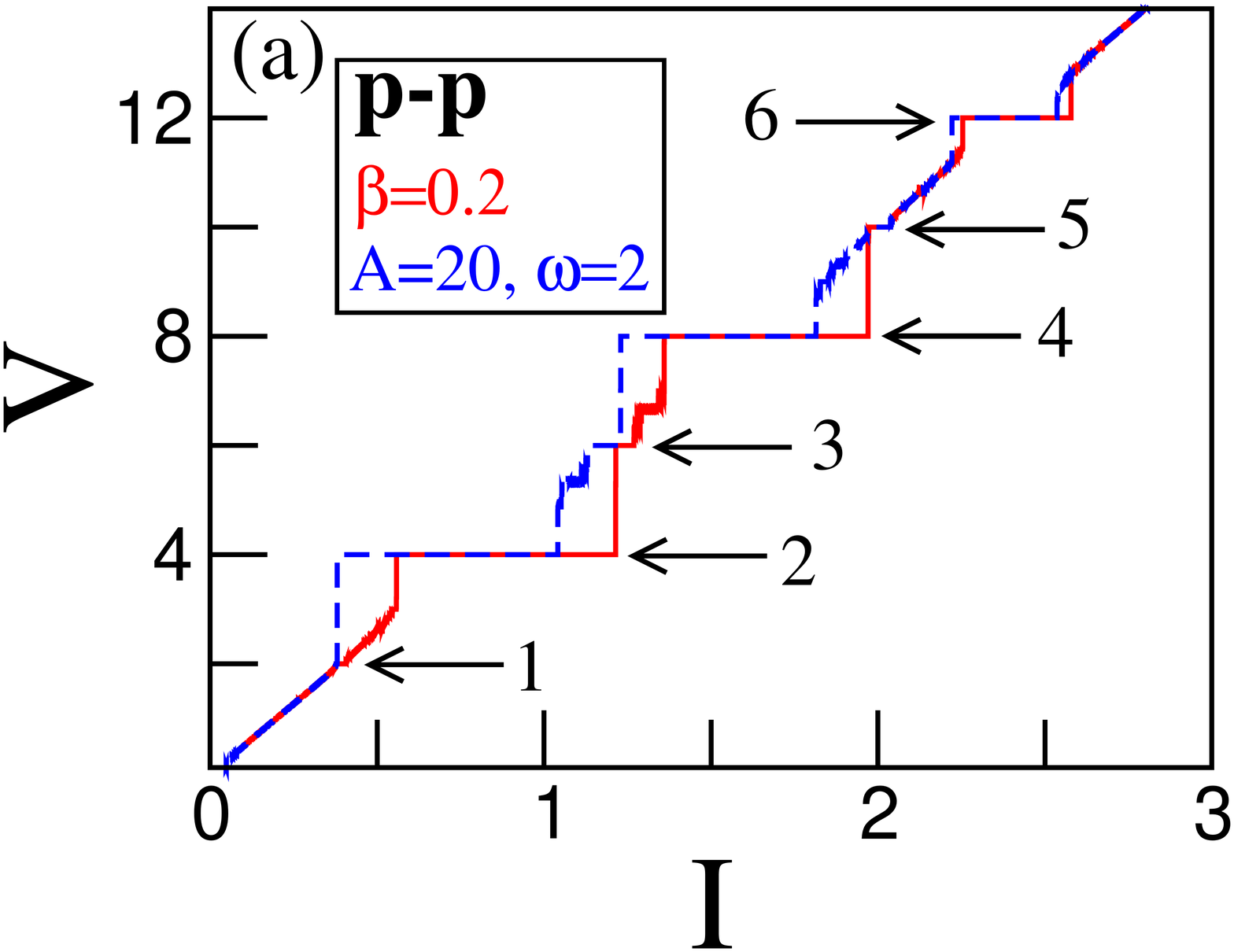}
\includegraphics[width=0.49\columnwidth]{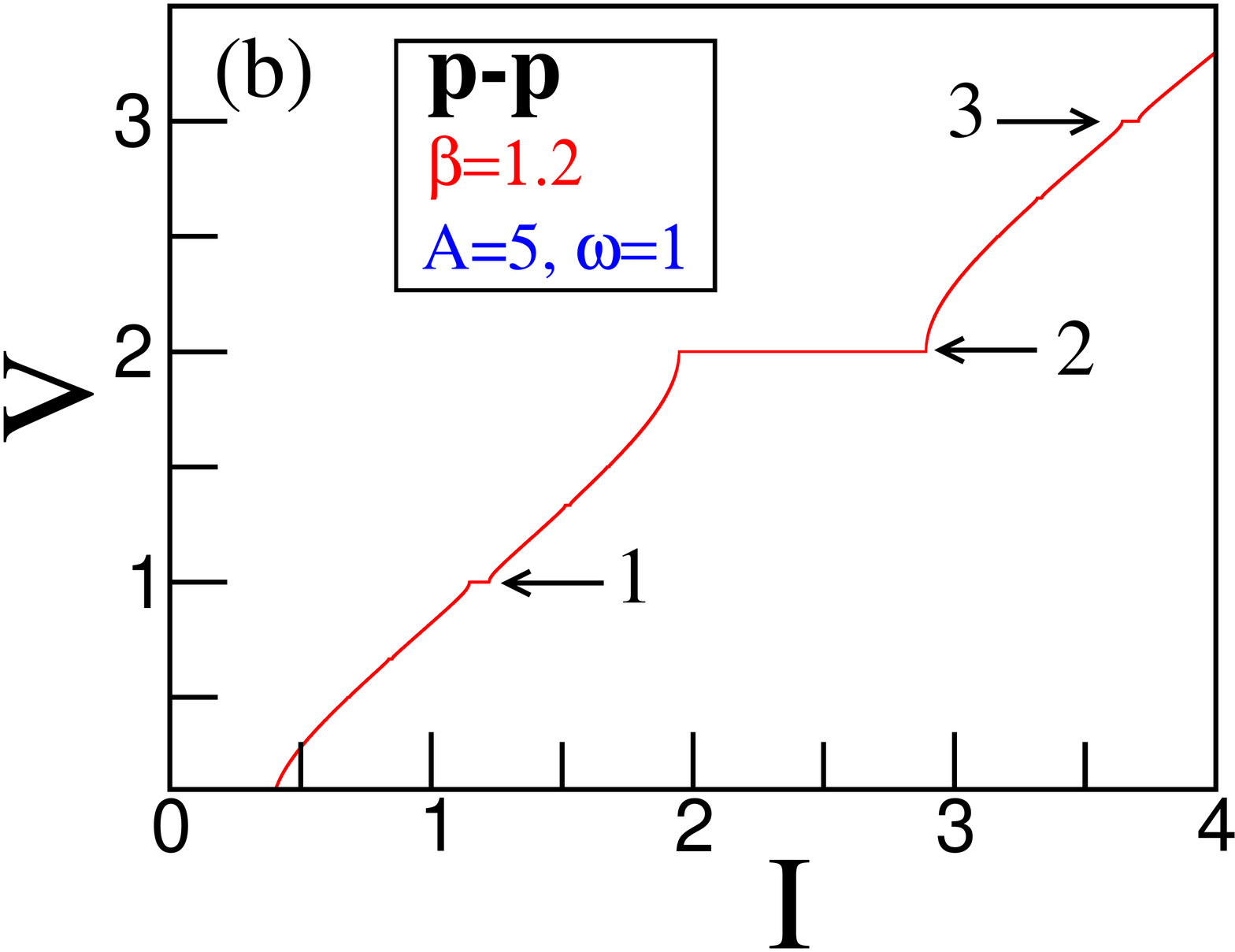}
\includegraphics[width=0.49\columnwidth]{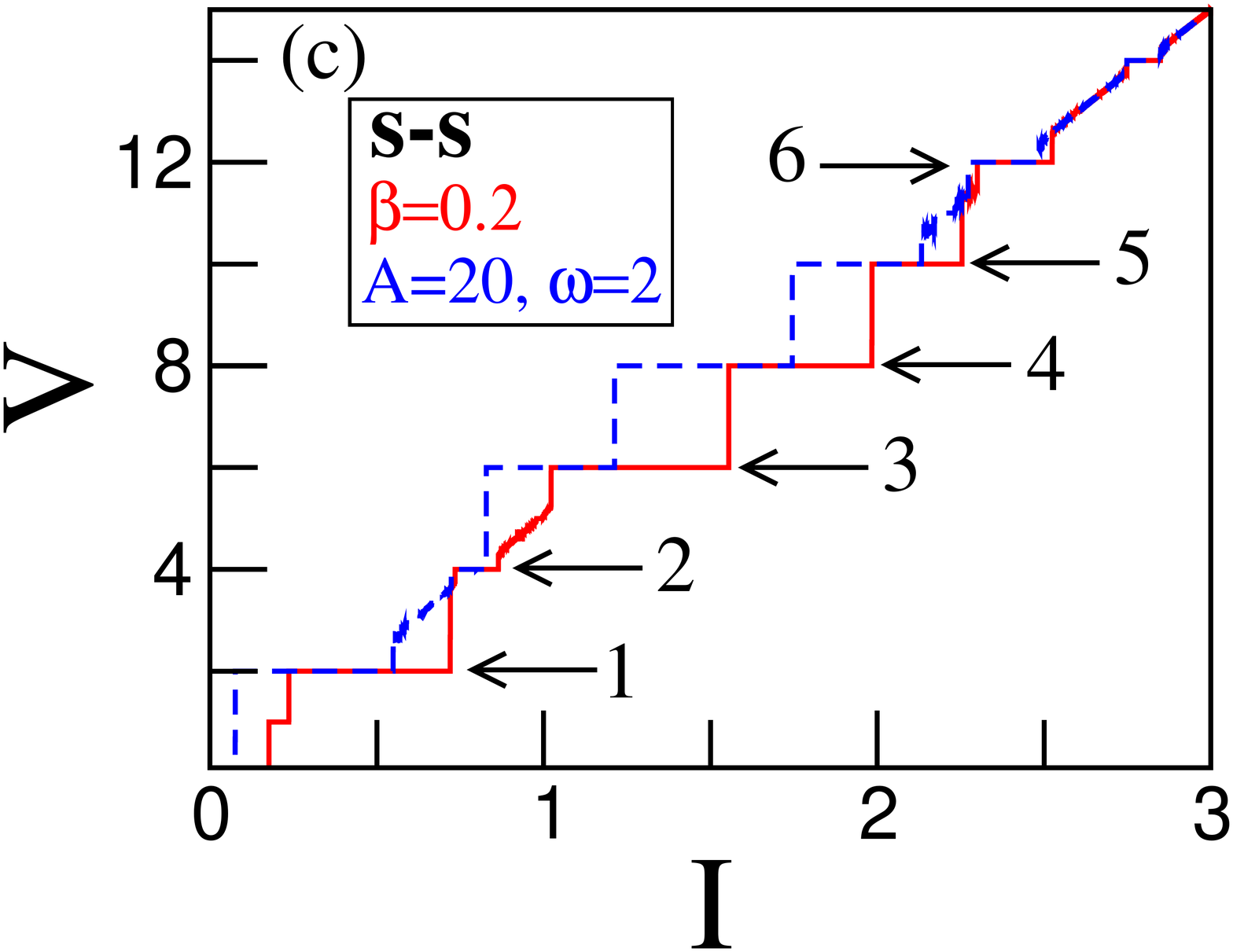}
\includegraphics[width=0.49\columnwidth]{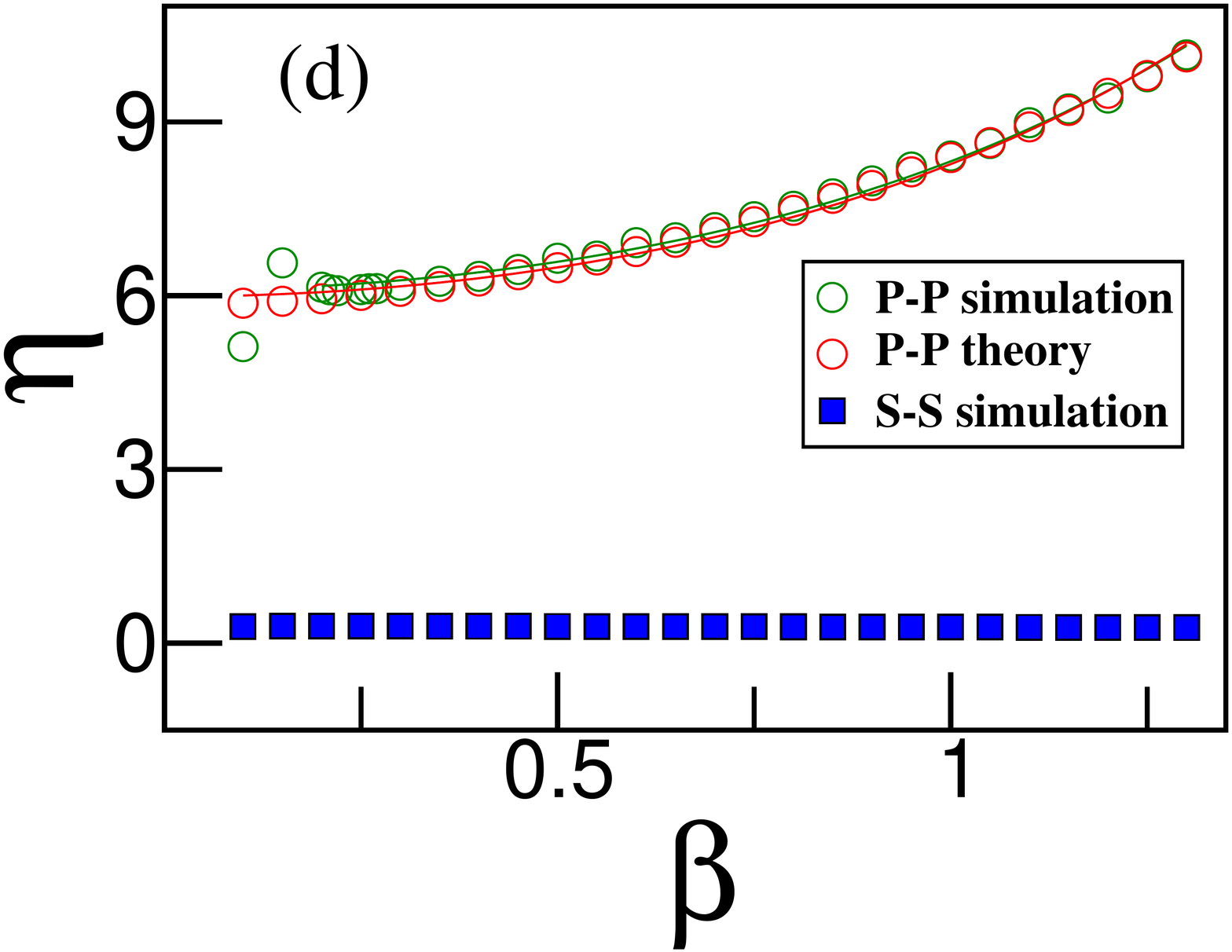}
\end{center}
\caption{(Color online) Top panels: CVC of $p$-wave Josephson
junction. Fig.\ref{fig2}a shows the CVC in the underdamped region
($\beta=0.2$) and for $A=20$ and $\omega=2$. while Fig.\ \ref{fig2}b
represents the overdamped region ($\beta=1.2$) with $A=5$ and
$\omega=1$ (in appropriate dimensionless units; see text). Bottom
Panels: Fig.\ \ref{fig2}c shows the CVC of $s$-wave junction (all
parameters are same as in Fig.\ \ref{fig2}a). Fig.\ \ref{fig2}d
shows the ratio of the widths of the even and odd Shapiro steps,
$\eta= W_{\rm even}(2\omega)/W_{\rm odd}(\omega)$ for the p-wave and
the s-wave junctions (inset) as a function of $\beta$ for $A=10$ and
$\omega=3$. For p-wave, $\eta \sim e^{0.31 \beta^2}$ while for
$s$-wave, $\eta$ does not vary with $\beta$. Figs.\
\ref{fig2}(a),\,(b), and\, (c) have $V$ and $I$ scaled in units of
$\hbar/(\tau e)$ and $I_c$ respectively. The red solid (blue dashed)
curves in panels (a) and (c) correspond to data for
increasing(decreasing) current sweeps; these data coincides in the
overdamped region as shown in panel (b). } \label{fig2}
\end{figure}

The central results that we obtain from this analysis are as
follows. First, for topological superconductors hosting Majorana
subgap states, we find that for a significant range of external
coupling and in the underdamped region $\beta < 1$, show both even
and odd Shapiro steps as expected from the analytical results
obtained in Sec.\ \ref{majan}. The even steps at $V=2n \hbar
\omega/e$ are enhanced compared to their odd counterparts at
$V=(2n+1)\hbar \omega/e$ as shown in Figs.\ \ref{fig2}a and
\ref{fig2}b for underdamped ($\beta <1$) and overdamped ($\beta >1$)
regions respectively. Fig.\ \ref{fig2}c shows an analogous plot for
the $s$-wave superconductors. The dominance of the even steps over
the odd ones is characterized by $\eta_1 \equiv \eta$ (Eq.\
\ref{etaan}). A plot of $\eta$ as a function of $\beta$ shown in
Fig.\ \ref{fig2}d, demonstrates that $\eta \sim \exp(0.3 \beta^2)$
for junctions with Majorana modes. We also note that the theoretical
result for $\eta$ obtained from Eq.\ \ref{etaan} provides a
near-perfect match with the exact numerics demonstrating the
accuracy of the analytical solution over a wide range of $\beta$. We
further note that the behavior of $\eta$ as a function of $\beta$ is
in complete contrast to its counterpart for $s$-wave superconductors
where $\eta$ does not vary appreciably with $\beta$ as shown in
Fig.\ \ref{fig2}d. Thus the exponential dependence of $\eta$ on the
junction capacitance $C$ constitutes a phase sensitive signature of
the presence of the Majorana modes. Note that it is generally
expected that only even Shapiro steps occur in I-V characteristics
of Josephson junctions which supports Majorana modes due to $4\pi$
periodicity of the Josephson current \cite{ks2,julia1,hassler1};
this is a consequence of analysis of the problem in the limit of
zero junction capacitance $C \to 0$ \cite{ks2,julia1,hassler1} where
$\beta \to \infty$ leading to vanishing of the odd steps. However,
our study show that underdamped RCSJ of such unconventional
Josephson junctions with finite $C$ can display both even and odd
steps. Thus the presence of odd Shapiro steps do not necessarily
signify the absence of Majorana modes specially if the RCSJ is
underdamped.

\begin{figure}[t!]
\begin{center}
\includegraphics[width=0.9\columnwidth]{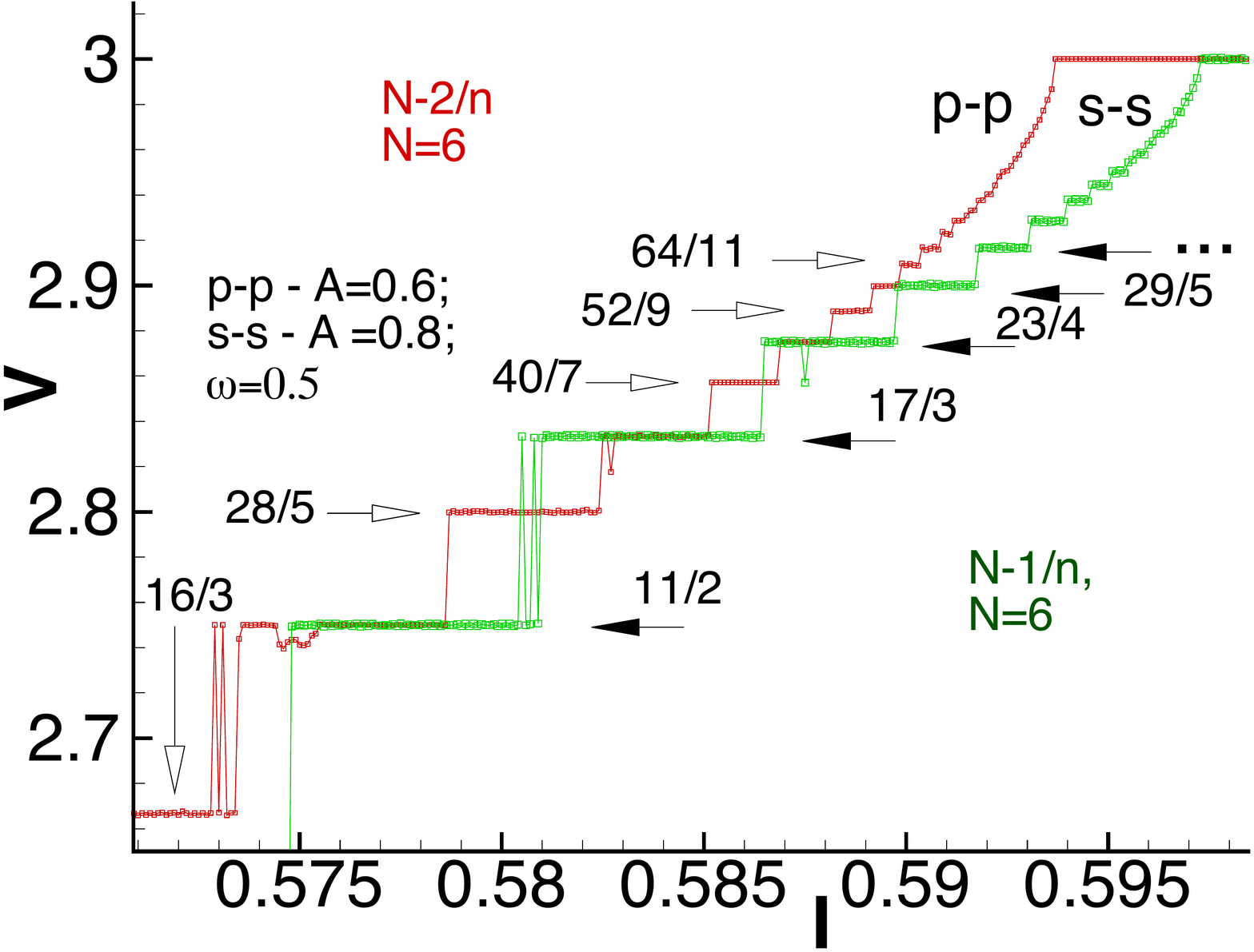}
\end{center}
\caption{(Color online) Plots of the self-similar structure for
$p$-wave ($A=0.6$) $s$-wave ($A=0.8$) Josephson junctions for
$D=0.7$, and $\omega=0.5$. These additional fractions marked with
arrows pointing to the right belong to the additional sequence
characteristics of Josephson junctions with Majorana subgap states
and obey Farey sum rule. $V$ and $I$ are scaled in units of $\hbar
/(e\tau)$ and $I_c$ respectively; see text for details.}
\label{fig3}
\end{figure}

\begin{figure}[t!]
\begin{center}
\includegraphics[width=0.9\columnwidth]{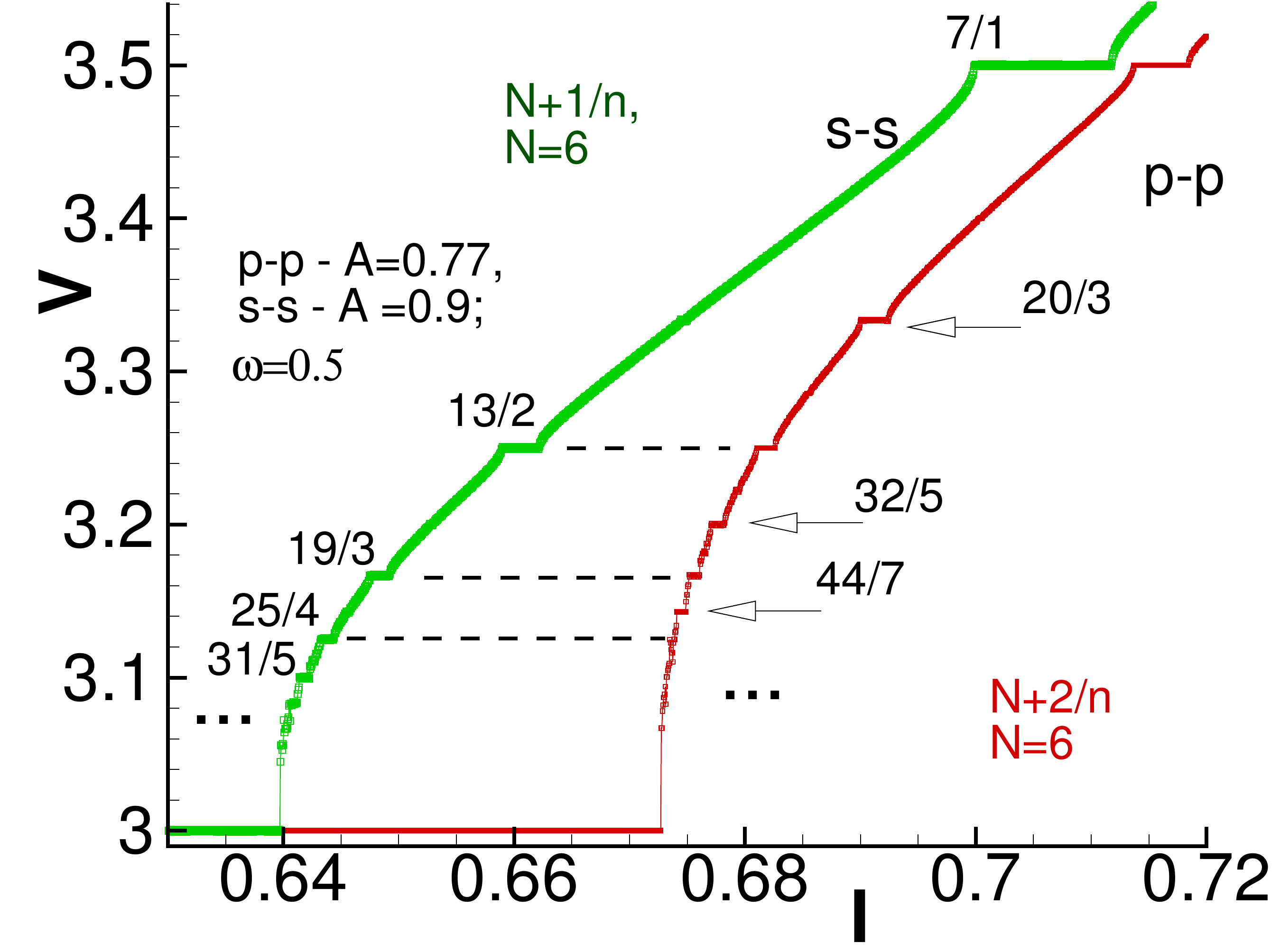}
\end{center}
\caption{(Color online) Same as in Fig.\ \ref{fig3} but with
different amplitudes of external radiation [ $p$-wave ($A=0.77$)
$s$-wave ($A=0.9$)]. Note that the additional sequence of steps for
$p-$ wave persists.} \label{fig4}
\end{figure}

Another qualitative difference between the I-V characteristics of
Josephson junctions hosting Majorana states with their conventional
counterparts occur in the devil staircase structures of the Shapiro
steps occurring within a fixed bias current intervals in these
junctions. Such steps are known to occur for conventional $s$-wave
junctions and their voltage-frequency relation can be represented by
a continued fraction as
\begin{eqnarray}
V &=& \left[N \pm 1/( n \pm (1/m \pm 1/(p \pm 1 ...)))\right]
\omega,
\end{eqnarray}
where $N,n,m,p$ are integers \cite{yury1}. The fractions involving
$N$ are termed as first level fractions, those with $N$ and $n$ are
second level fractions, and so on. To distinguish between $s$-wave
and $p-$wave junctions, we investigate the structure of these steps
for both $p-$ and $s-$wave junctions as shown in Figs.\ \ref{fig3}
and \ref{fig4}. We find that for $s$-wave, only second level
subharmonics are present and $V=6\omega$ is approached from below
with steps occurring at $V= N(1-1/n)\omega$ for different $N$ and
$n$. In contrast, for $p$-wave junctions, one finds that in addition
to the steps which occur for the $s-$wave junctions, there are
additional steps corresponding to $V= [(N \mp 2)][1+1/(n\pm
1/m)]\omega$ (Figs.\ \ref{fig3} and \ref{fig4}). This difference in
structure lead us to hypothesize that in contrast to $s-$wave
Josephson junctions, the continued fractions for $p-$wave shows
additional fractions all of which are consistent with the Farey sum
rule; {\it i.e.}, the widest phase-locked region (the widest step)
between any two resonances $p/q$ and $m/n$ for these fractions is
given by $(p+m)/(q+n)$. We therefore suggest that the presence of
these additional specific continued fractions may be considered as a
signature of Majorana modes.

Two specific examples of this phenomenon is presented in Fig.\
\ref{fig3} and \ref{fig4}. In Fig.\ \ref{fig3}, we find that the
continued fraction $V= (N-1/n)\omega$ with $N=6$ which appears in
$s-$wave Josephson junctions at $\beta=0.2$, $\omega=0.5$ and
$A=0.8$, is also appeared in the $p-$wave junction at smaller
amplitude of radiation $A=0.6$. However, as clearly demonstrated in
Fig.\ \ref{fig3}, the steps for the $p-$ wave junction constitutes
an additional sequence of continued fraction $V=
[(N-2)][1+1/(n+1/m)]\omega$ with $N=6$, $n=2$ and for several $m$.
Further, as shown in Ref.\ \onlinecite{yury1}, the continued
fraction $V= (N+1/n)\omega$ with $N=6$ which appears in $s-$wave
Josephson junctions at $A=0.9$. In contrast, for $p-$wave Josephson
junctions, as shown in Fig.\ \ref{fig4}, there is an additional
continued fraction $V= [(N+2)/2][1+1/(n-1/m)]\omega$ with $N=6$ and
$n=2$. In both cases, additional steps occur when Majorana subgap
states are present; thus, although we have not been able to find a
precise analytical expression relating the $4 \pi$ periodicity of
$I_J(\phi)$ and these additional steps, we conjecture that these
additional steps are a consequence of the subgap Majorana bound
states of the superconductors forming the junctions.

\section{ Junctions with Dirac Fermions}
\label{dir1}

For superconducting junctions hosting Dirac quasiparticles such as
in graphene \cite{been1,ks1}, the pair-potential has $s$-wave
symmetry: $\Delta_L= \Delta_0$ and $\Delta_R= \Delta_0 \exp(i
\phi)$. Here pairing occurs between electrons with opposite spin and
momenta; in graphene this necessitates pairing between electrons of
$K$ and $K'$ valleys \cite{grrev}. This the BdG equations are
described in terms of $4$ component wavefunctions $\psi = (\psi_{A
\uparrow}^K, \psi_{ B \uparrow}^K, \psi_{A \downarrow}^{K' \dagger},
\psi_{B \downarrow}^{K' \dagger})$ and is given by
\begin{eqnarray}
[(H'-V(x))\tau_3 + (\Delta (x) \tau_{+} +{\rm h.c.}) ]\psi = E \psi,
\label{bdg2}
\end{eqnarray}
where $A,B$ denotes sublattice indices $H'_{\beta} = \hbar v_F (-i
\sigma_x \partial_x \pm \sigma_y k_y) -\mu$, $\vec \tau$ and $\vec
\sigma$ denote Pauli matrices in valley and pseudospin (sublattice)
spaces respectively, $\mu$ is the Fermi energy, $v_F$ is the Fermi
velocity of the Dirac quasiparticles described by $H'$ and $+(-)$
sign corresponds to electrons in $K(K')$ valley. The pair-potential
takes the form $\Delta(x) = \Delta_L \theta(d/2+x) + \Delta_R
\theta(x-d/2)$ and $V(x)= V_0 \theta(d/2-x)\theta(d/2+x)$. The
localized subgap Andreev states is obtained by demanding the
continuity of $\psi$ at $x=\pm d/2$ and is given, in the thin
barrier limit ($ V_0 \to \infty, \, d\to 0, \, {\rm and}\, V_0
d/\hbar v_F = \chi_2$) by \cite{ks4}
\begin{eqnarray}
E_2 &=& \pm \Delta_0 \sqrt{1- T(k_y,\chi_2) \sin^{2}(\phi/2)}
\label{endir}
\end{eqnarray}
where $T(k_y, \chi_2) =
\cos^2(\gamma)/(1-\cos^2(\gamma)\sin^2(\chi_2))$ is a measure of
transparency of the junction and $\sin(\gamma)= \hbar v_F k_y/\mu$.
Note that Eq.\ \ref{endir}, in contrast to Eq.\ \ref{enmaj}, is $2
\pi$ periodic in $\phi$. The corresponding Josephson current at zero
temperature is given by $I_2(\phi)= \frac{2e}{\hbar}
\int_{-\pi/2}^{\pi/2} d\gamma \cos(\gamma) \partial E_2/\partial
\phi$ and leads to
\begin{eqnarray}
I_2 &=& I_0 \Delta_0 \int_{-\pi/2}^{\pi/2} d \gamma \cos(\gamma)
\sin(\phi) T(\gamma, \chi_2)/|E_2| \label{jos2}
\end{eqnarray}
where $I_0= e\Delta_0 E_F L/(2 \hbar^2 \pi v_F)$. Eq.\ \ref{jos2}
shows that the Josephson current is an oscillatory function of the
dimensionless barrier strength for such junctions. We note that
whereas the form of Eqs.\ \ref{endir} and \ref{jos2} are generic for
$s-$wave conventional tunnel junctions,  the oscillatory dependence
of $T$ on $\chi$ is a consequence of Dirac nature of graphene
quasiparticles and is not observed in junctions made of conventional
superconductors \cite{ks4}.

\begin{figure}[t!]
\begin{center}
\includegraphics[width=0.475\columnwidth]{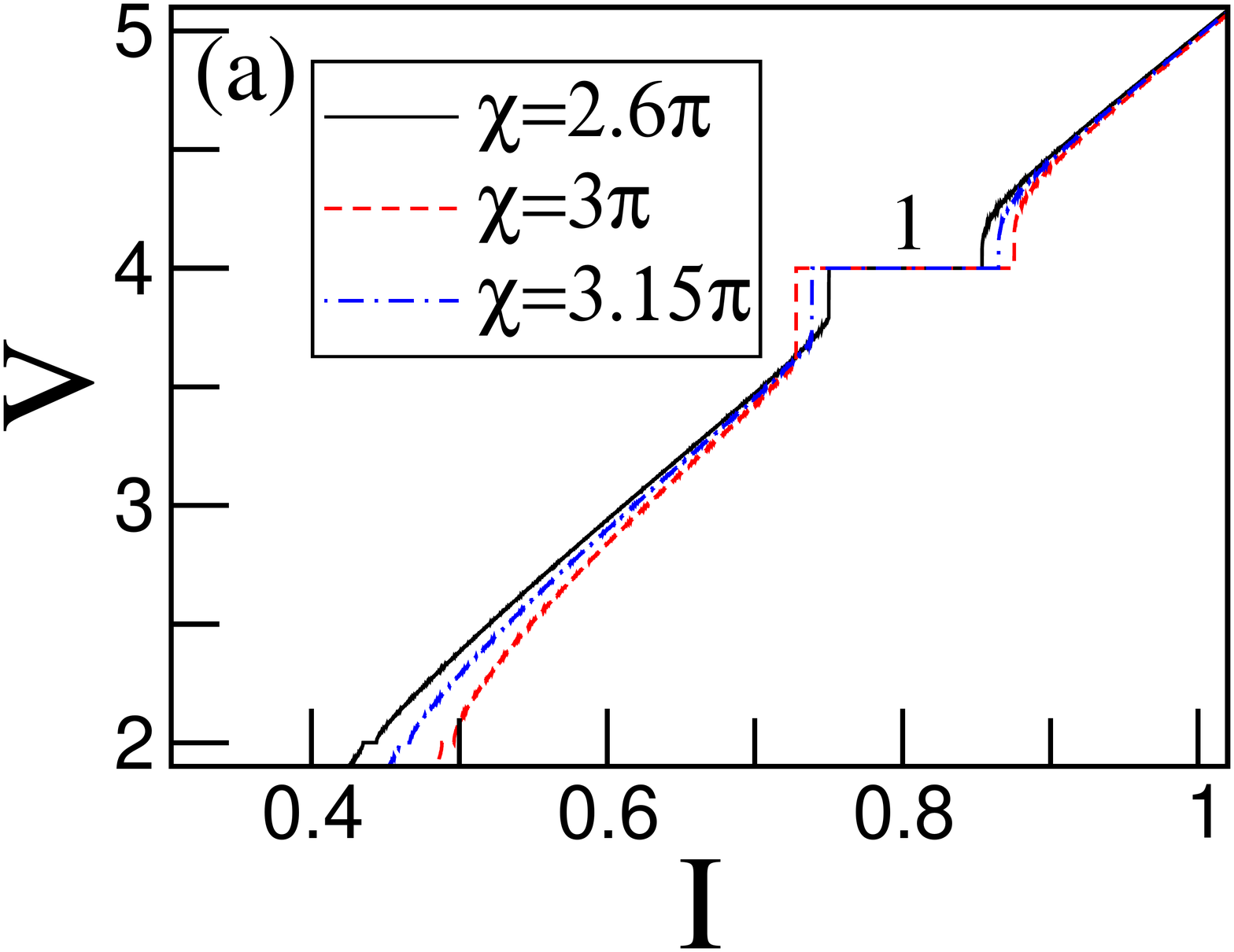}
\includegraphics[width=0.49\columnwidth]{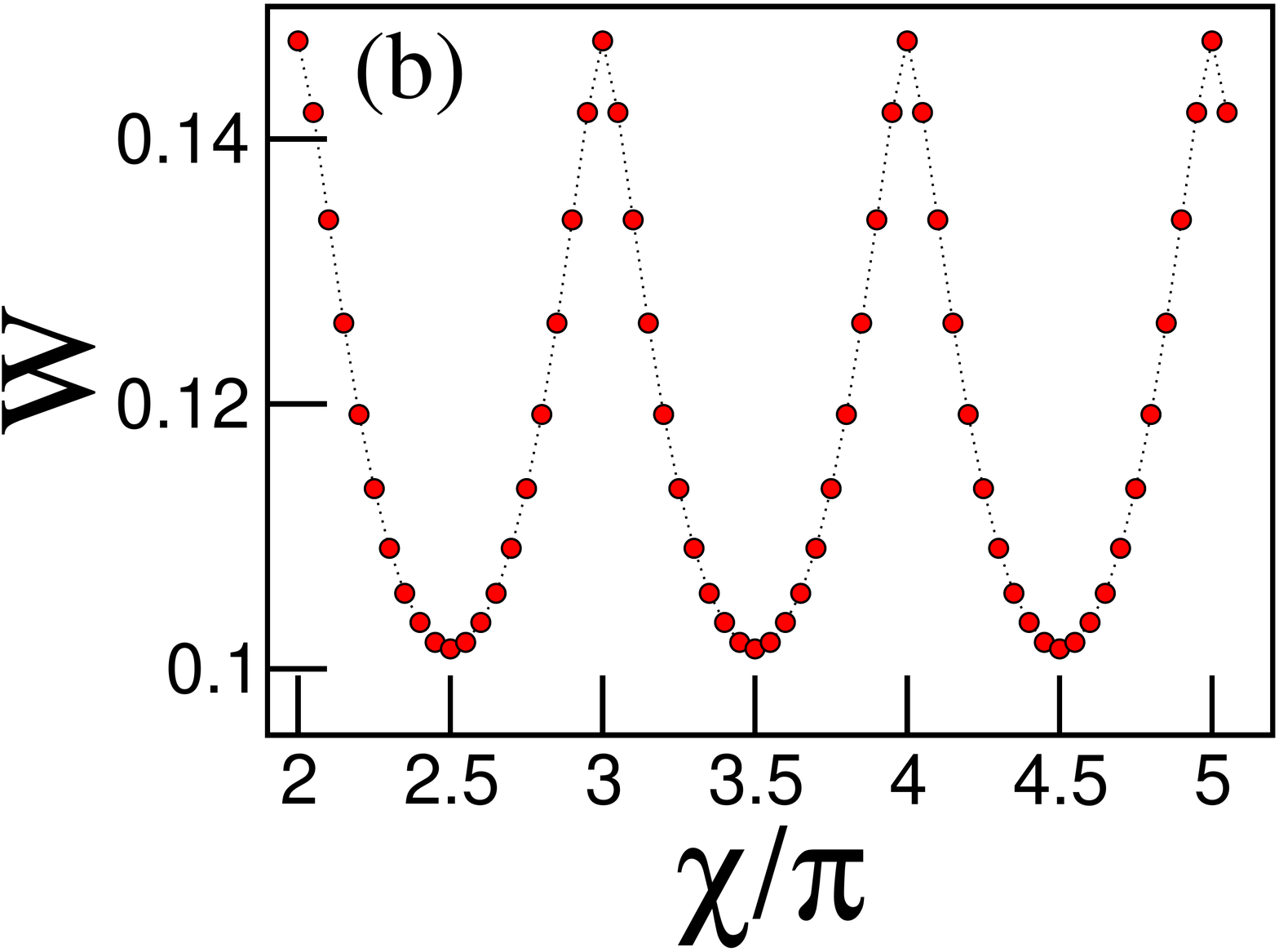}
\end{center}
\caption{(Color online) (a) CVC of graphene Josephson junction with
$A=1$ and $\omega=4$ for several values of $\chi$ indicating the
variation in the width of the main Shapiro step at $\omega$. (b)
Plot of the width of the Shapiro step at $\omega$ as a function of
$\chi$ showing $\pi$ periodic oscillatory behavior.}\label{fig5}
\end{figure}

To chart out the I-V characteristics of the superconducting
junctions which host such Dirac quasiparticles, we analyze Eq.\
\ref{rcsj1} numerically with $I_J=I_2$ and obtain the corresponding
Shapiro step structure. The procedure followed here is identical to
the one charted out in Sec.\ \ref{majnu}. We find that the Shapiro
step structure in the I-V characteristics is same as the
conventional $s$-wave superconductor displaying harmonic odd and
even steps as shown in Fig.\ \ref{fig5}a. However, the width of
these steps, $W$, vary with the dimensionless barrier potential
$\chi_2$ in an oscillatory manner as shown in Fig.\ \ref{fig5}b.
This is in complete contrast with the dependence of $W$ in the
conventional junctions where the step widths are monotonically
decreasing function of the barrier potential. This behavior of $W$
can be qualitatively understood as follows. In a RCSJ, $W$, can be
related to the magnitude of $I_J$ which, in turn, depends on the
transparency of the junction: $W \sim (1+\chi_1^2/4)^{-1/2}$ for
conventional junctions. In conventional junction, increase of the
barrier potential $\chi_1$ leads to a monotonic decrease of the
transparency; consequently, $W$ decrease monotonically with
increasing $\chi_1$. However, for a RCSJ made out of Dirac
materials, the transparency of the junction, $T(k_y,\chi_2)$, is an
$\pi$ periodic oscillatory function of the dimensionless barrier
strength $\chi_2$ with maxima at $\chi=n \pi$ due to the
transmission resonance condition of the Dirac quasiparticles
\cite{ks3}. Consequently, one expects $I_J$ and hence $W$ to
oscillate with $\chi_2$. This expectation is corroborated in Fig.\
\ref{fig5}b, where the $\pi$ periodic oscillation of the step width
is plotted as a function of $\chi_2$. We note that such a
oscillatory behavior is a direct manifestation of the transmission
resonance condition of the Dirac quasiparticles; it thus provides a
qualitative distinction between Josephson junction hosting
Schrodinger and Dirac quasiparticles.

\section{Discussion}
\label{conc1}

In this work we have studied the I-V characteristics of a RCSJ where
the superconductors making up the junction either hosts subgap
Majorana bound states or have Dirac-like character of the Bogoliubov
quasiparticles. The former set of junctions occur for $p$-wave
superconductors \cite{ks1} or 1D nanowires \cite{sup1,sup2} with
strong spin-orbit coupling and transverse magnetic field while
graphene superconduction junctions provide an example of the latter
class. We find that the I-V characteristics of RCSJs for each of
these classes of junctions are qualitatively different from their
conventional counterparts. Thus such junctions may serve as phase
sensitive detectors of Majorana and Dirac fermions realized using
superconducting platforms.

For junctions hosting subgap Majorana states, we find two essential
characteristics which are qualitatively different from their
$s$-wave counterparts. First, the odd Shapiro steps are subharmonic
in nature; the ratio of their width with that of adjacent even
Shapiro is a decreasing function of the junction capacitance. This
is in contrast to the conventional $s-$wave junctions where the
ratio is largely independent of $C$. We note that our result in this
regard shows that the absence of odd Shapiro steps is a sufficient
condition for having subgap Majorana modes; however it is not
necessary since such an absence requires, in addition to the
presence of the Majorana modes, resistive Josephson junctions. Our
result thus constitute a generalization of detection criteria for
Majorana modes realized using superconducting platform. Second, we
find that the devil staircase structure of the Shapiro steps in
Josephson junctions with Majorana subgap states involves additional
sequences which satisfies Farey sum rule. This feature, as shown in
our work, constitutes a qualitative difference between Josephson
junctions with and without Majorana subgap states.

For junctions with Dirac quasiparticles, we find that even with
$s$-wave symmetry, the Shapiro step width is a $\pi$ periodic
oscillatory function of the barrier potential of the junction. We
trace the origin of this phenomenon to the transmission resonance of
the Dirac-Bogoliubov quasiparticles in such superconductors and
demonstrates that the oscillatory behavior is a qualitatively
distinct signature of Dirac nature of the superconducting
quasiparticles.

The numerical estimate of typical frequencies at which the devil
staircase structure can be obtained as follows. For standard
experiments $ I_c \sim 1$nA and $ C_0 \simeq 1$pF. Using these
numbers one can estimate, $\omega_p = \sqrt{2e I_c/(\hbar C_0)}
\simeq 1$GHz. In all the plots, that we have used $\omega$ ranges
between $0.5 \omega_p \simeq 0.5$GHz to $ 2 \omega_p \simeq 2$GHz.
In particular, the devil staircase structure is seen at an external
radiation frequency of $0.5$GHz. The self-similar structure is seen
at energy range of $5-6 \omega$ which is around $2-3$GHz. In this
context, we note that the required frequency range is large small
enough to avoid possible smearing due to $2 \pi$ periodicity arising
due to multimode effects \cite{hassler1}. In addition, this estimate
holds for zero-temperature analysis; however it is expected to be
qualitatively accurate for $k_B T \ll \Delta_0$ when quasiparticle
poisoning and thermal decoherence rates do not play a significant
role. Also, we note that it is possible to model the dissipation and
noise in the junction by assuming it to be coupled through a thermal
bath using the standard Caldeira-Leggett formalism \cite{cal1}; the
Langevin or saddle point equation corresponding to that analysis at
low temperature, where the effects of white noise can be ignored,
reduces to Eq.\ \ref{rcsj1} of our work with the resistive term
being renormalized by the coefficient of dissipation. This formalism
also allows for study of effect of quantum fluctuations and noise in
such junctions beyond saddle point approximation which is left as a
possible subject of future study.

The experiments to test our theory involves on measurement on RCSJ
under an applied radiation with definite amplitude $A$ and frequency
$\omega$. Such experiments are rather standard for $s-$wave
junctions \cite{swaveexp}; more recently, such experiments have been
performed for 1D Majorana nanowire setup with resistive junctions
\cite{exp2}. Our specific suggestion involves measurement of $\eta$
as a function of the effective junction capacitance for JJS with
Majorana bound states in a circuit with finite capacitance which can
be modeled by a RCSJ; we predict the presence of subharmonic odd
Shapiro steps for such junctions whose width depends on the junction
capacitance and lead to an exponential dependence of $\eta$ with the
junction capacitance (See Fig.\ \ref{fig2}d). In addition, we
suggest the presence of an additional sequence in the devil
staircase structure of the Shapiro steps. For junctions with Dirac
quasiparticles, which can be made with graphene \cite{grapheneexp},
we predict that the width of the Shapiro steps will display $\pi$
periodic oscillatory dependence with the junction barrier potential.

In conclusion, we have studied RCSJ Josephson junction circuits and
have shown that they can serve as phase sensitive detectors for both
Majorana and Dirac quasiparticles in such junctions. We have charted
out the properties of such junctions which are qualitatively
distinct from their $s$-wave counterparts and have suggested
experiments which can test our theory.


\begin{thebibliography}{99}
\bibitem{comment1} The energy dispersion remain same for 1D nanowire
Josephson junctions  with effective $p$-wave pairing
\cite{sup1,sup2} without the transverse momentum dependence of
$\chi_1$. All our results are therefore applicable to such junctions
sans the integration over transverse momentum modes in Eq.\
\ref{jos1}.
\end{thebibliography}
\end{document}